\documentclass[a4paper,fleqn,usenatbib,useAMS]{mn2e}
\usepackage{graphicx}
\usepackage{amsmath}
\usepackage{amssymb}
\usepackage{multicol}
\usepackage{bm}
\usepackage{aas_macros}

\usepackage[T1]{fontenc}
\usepackage{ae,aecompl}

\usepackage{color}
\usepackage[varg]{txfonts}

\title[Circumbinary discs around merging stellar--mass black holes]{Circumbinary discs around merging stellar--mass black holes}

\author[Martin et al.]{Rebecca
  G. Martin$^1$\thanks{rebecca.martin@unlv.edu}, Chris Nixon$^2$, Fu-Guo Xie$^{3,2}$
  and Andrew King$^{2,4,5}$ 
\\$^1$Department of Physics and Astronomy, University
  of Nevada, Las Vegas, 4505 South Maryland Parkway, Las Vegas, NV
  89154, USA 
\\$^2$Department of Physics and Astronomy, University of Leicester, University Road, Leicester LE1 7RH, UK 
\\$^3$Key Laboratory for Research in Galaxies and Cosmology, Shanghai Astronomical Observatory, Chinese Academy of Sciences, 80 Nandan Road, \\ Shanghai 200030, China 
\\$^4$Anton Pannekoek Institute, University of Amsterdam, Science Park 904, 1098 XH Amsterdam, Netherlands 
\\$^5$Leiden Observatory, Leiden University, Niels Bohrweg 2, NL-2333 CA Leiden, Netherlands}

\date{}
\pagerange{\pageref{firstpage}--\pageref{lastpage}} 
\pubyear{2018}
 
\begin{document}
\maketitle
\label{firstpage}

\begin{abstract}
A circumbinary disc around a pair of merging stellar--mass black
holes may be shocked and heated during the recoil of the merged hole,
causing a near-simultaneous electromagnetic counterpart to the
gravitational wave event. The shocks occur around the recoil radius,
where the disc orbital velocity is equal to the recoil velocity. The
amount of mass present near this radius at the time of the merger is
critical in determining how much radiation is released. We explore the
evolution of a circumbinary disc in two limits. First, we consider an
accretion disc that feels no torque from the binary. The disc does not
survive until the merger unless there is a dead zone, a region of low
turbulence. Even with the dead zone, the surface density in this case
may be small. Second, we consider a disc that feels a strong binary
torque that prevents accretion on to the binary. In this case there is
significantly more mass in regions of interest at the time of the
merger. A dead zone in this disc increases the mass close to the
recoil radius. For typical binary-disc parameters we expect accretion
to be significantly slowed by the resonant torque from the binary, and
for a dead zone to be present. We conclude that provided significant
mass orbits the binary after the formation of the black hole binary
and that the radiation produced in recoil shocks can escape the flow
efficiently, there is likely to be an observable electromagnetic
signal from black hole binary mergers.
\end{abstract}


\section{Introduction}
\label{intro}

The direct detection of gravitational waves (GW) from binary black
hole mergers has naturally raised the question of whether there could
be a corresponding electromagnetic (EM) signal.  Recently
\cite{deMink2017} suggested that the gravitational effect of a BH + BH
merger on a circumbinary disc might lead to an EM signal delayed by
the light--travel time from the merging binary to this disc \citep[see
  also][where this effect is applied to mergers of supermassive black
  holes]{Rossietal2010}. Given the lack of simultaneity, identifying
this latter signal would require spatial coincidence, and so
positional information from several GW detectors. This should be
available as new detectors become operational.

At least two detailed binary evolution scenarios, the classical
common--envelope and chemically homogeneous pictures, show how a
massive binary can remain bound as it evolves to a BH + BH binary (see
the discussion in de Mink \& King 2017, Section 2, and references
therein). In both cases the two stars shed large fractions of their
original masses. Unless all of this mass is expelled to infinity, or
somehow completely re--accreted by the black holes, a residual
circumbinary disc must remain, as the only stable configuration for the
shed mass.

In this paper we consider the evolution of this circumbinary
disc around a merging black hole binary. Specifically, we are
interested in whether there is sufficient material in the disc at the
time of the merger for an electromagnetic signal to be generated by
the perturbation from the gravitational wave recoil and the sudden
drop in mass of the binary.

The circularisation radius of material falling back after a supernova
explosion is very small \citep{Perna2014, Perna2016}. This suggests
that the angular momentum of the ejected material is small in the
frame of the star. Thus the material is ejected with a velocity equal
to that of the mass losing star plus a velocity radial from the mass
losing star with magnitude $v_{\rm eject}$. For ejection velocities
greater than the escape velocity from the binary, $v_{\rm eject} >
v_{\rm esc}$, the material is lost. For $v_{\rm eject} \ll v_1$, where
$v_1$ is the orbital velocity of the mass--losing star, the material
does not escape the Roche-Lobe of the star. For $v_{\rm eject} \sim
v_1$ the material has a spread of angular momentum around the
angular momentum of the mass--losing star: the material at the back of
the star has less, and the material at the front of the star
has more. Thus we expect approximately half of the ejected
material to have a velocity such that it ends up in a circumbinary configuration
with a spread of angular momenta and thus circularisation
radii. Whether the material circularises, or passes within the binary
and cancels angular momentum with gas on similar but opposed orbits, to
circularise at smaller radii, requires more detailed numerical simulations than
we attempt here.

The angular momentum of the material forming the circumbinary disc
may not necessarily be initially aligned to the binary angular
momentum.  For a circular black hole binary, differential
precession driven by the binary torque combined with viscous
dissipation causes the disc to move towards alignment or counter--alignment 
with the binary depending only on the initial inclination
\citep[e.g.][]{Kingetal2005,Nixonetal2011b,Foucart2013,Foucart2014}. As
the disc angular momentum is usually significantly smaller than the
binary angular momentum \citep{Nixon2012}, the alignment or
counteralignment condition is straightforward: if the disc begins
closer to alignment, it aligns, and if it begins closer to
counteralignment it counteraligns. How this alignment proceeds depends
upon properties of the disc. \cite{PP1983} showed that the propagation
of warps occurs in two distinct ways, the dominant mechanism
determined by the relative magnitudes of the disc viscosity and
pressure. If the \cite{SS1973} viscosity parameter is less than the
disc aspect ratio, $\alpha<H/R$, then the warp propagates through
waves, otherwise the warp evolves diffusively \citep[see][for a review
of warped disc physics]{Nixon2016}. Since the disc aspect ratio of
circumbinary discs around black holes is likely to be small, the inner
parts of the disc are likely to be aligned (or counteraligned) while
the outer parts of the disc may be strongly warped.

The strength of the tidal torque from the binary on the inner parts of the
disc depends upon the inclination of the angular momentum of the material,
being strongest when it is aligned to the binary angular momentum
\citep[e.g.][]{Lubowetal2015,Miranda2015}. For a retrograde
circumbinary disc around a circular binary the tidal torque from Lindblad
resonances is zero \citep{Nixonetal2011a}. \cite{Nixon2015} showed that
resonances do occur in retrograde circumbinary discs around eccentric
binaries, but that they are weak enough that they only modulate the accretion
flow on to the binary rather than providing a torque of sufficient strength to
arrest accretion.

The accretion rate from the inner edge of the circumbinary disc on to the
components of a binary depends upon the strength of the tidal torque and
properties of the disc. This has received significant attention as it impacts
a variety of astrophysical systems \citep[e.g.][]{Artymowicz1994,
  Artymowicz1996,Roedig2012,Shi2012,DOrazio2013,Farris2014}. The gas is
subject to two competing torques. The binary transfers angular momentum to the
gas, causing the inner disc to move outwards, and viscosity redistributes that
angular momentum to larger radii in the disc, allowing the inner disc to
shrink. Thus discs subject to a weak torque (or strong viscosity) can
move in and accrete on to the binary components. In contrast a strong torque
(or weak viscosity) leads to the disc being held out. For the small disc
aspect ratio expected in black hole binaries, the accretion rate is
significantly suppressed \citep{Ragusa2016}. If the tidal torque is
sufficiently strong, accretion on to the binary may be completely halted.
Disc solutions in this case correspond to {\it decretion} discs
\citep{Pringle1991} rather than traditional {\it accretion} discs in which
material flows freely through the inner boundary \citep{Pringle1981}.

In Section~\ref{em} we first consider how much material is required in
a circumbinary disc for an observable electromagnetic signal to be
generated. In the rest of this work, we then estimate how much
material is present in an evolving circumbinary disc.  Because of the
uncertainties in the strength of the tidal torque and the resulting
accretion flow on to the binary, we consider two extreme limits. In
Section~\ref{full} we consider the disc evolution in the case that the
binary provides no torque on the disc and material freely flows
inwards and is accreted on to the binary components. The disc is a
traditional accretion disc. In Section~\ref{zero} we examine the case
where the binary torque is strong enough to prevent all flow on to the
binary.  We draw our conclusions in Section~\ref{conc}.

\section{Electromagnetic signal from a circumbinary disc}
\label{em}

Gravitational wave emission during a binary black hole coalescence
produces a sudden drop in the total central mass. The 
weaker potential causes circumbinary disc orbits to expand
\citep[e.g.][]{Rosottietal2012}. This excites density waves which
dissipate their energy, giving an electromagnetic
signal. However, as the black holes merge the merged hole suffers a
momentum kick due to the asymmetry of the gravitational wave
emission. As the hole is kicked, the specific angular momentum of the
gas in the circumbinary disc is altered, leading the gas orbits to
become eccentric and causing shocks, dissipation and circularisation on
the local orbital timescale. This can give a rather stronger electromagnetic signal
\citep{Rossietal2010}.

We consider a black hole binary system of total mass $M$, composed of
two black holes of mass $M_{\rm bh}=30\,\rm M_\odot$. 
Heating by shocks occurs on the dynamical timescale
\begin{equation}
t_{\rm dyn}=\frac{GM}{v_{\rm rec}^3}=2.2\left(\frac{M}{60\,\rm M_\odot}\right)
\left(\frac{v_{\rm rec}}{1000\,\rm km\,s^{-1}}\right)^{-3}
\,\rm hr,
\end{equation}
where $v_{\rm rec}$ is the recoil velocity imparted to the centre of mass. 
The gas in the disc orbits the central mass at Keplerian angular
frequency $\Omega=\sqrt{G M/R^3}$, where $R$ is the radius from the
binary centre of mass.  The recoil radius is where the
Keplerian velocity of the disc is equal to the recoil velocity 
\begin{equation}
R_{\rm rec}=\frac{GM}{v_{\rm rec}^2}=11.5 \left(\frac{M}{60\,\rm
  M_\odot}\right) \left(\frac{v_{\rm rec}}{1000\,\rm
  km\,s^{-1}}\right)^{-2} \,\rm R_\odot
\end{equation}
\citep{deMink2017}.  The magnitude of the recoil velocity depends
sensitively on the pre--merger black hole spin vectors
\citep[e.g.][]{Blecha2016}. If the spins are within a few degrees of
alignment then the kick may be small, $\lesssim 600\,\rm km\,
s^{-1}$. However, if the spins are not aligned larger kicks up to
several thousand $\rm km\,s^{-1}$ can occur
\citep{Campanelli2007,Lousto2011}. Spin alignment is unlikely in X-ray
binaries
\citep[e.g.][]{Martinetal2007,Martinetal2009,Martinetal2008,MartinReisPringle2008,King2016}
or in SMBH binaries \citep{LG2013}, but could perhaps be facilitated
in close binaries where the stellar spins have been synchronised with
the binary orbit. In this work, we assume that kick velocities of
order $500-1000$ $\rm km\,s^{-1}$ are possible, and that some ejected
material settles into a circumbinary disc. We are interested in how
much material may still be present at the recoil radius at the time of
the merger.

The luminosity generated by the kick is calculated with rate of
  dissipation of kinetic energy as
\begin{equation}
L=\frac{fM_{\rm rec}v_{\rm rec}^2}{t_{\rm dyn}},
\end{equation}
where $M_{\rm rec}$ is an estimate of the mass of the disc within the
region that feels the kick strongly and $f$ is a scaling factor. The
scaling factor is calibrated to the numerical simulations by
\cite{Rossietal2010} who found $f\approx 0.1$ for a kick angle
$\theta=15^\circ$. The peak luminosity is somewhat insensitive to the
kick angle.  For typical parameters the luminosity is
\begin{equation}
L=f \frac{M_{\rm rec}v_{\rm rec}^2}{t_{\rm dyn}}=1.5\times 10^{43}\,
\left(\frac{f}{0.1}\right)
\left(\frac{M_{\rm rec}}{0.001\,M}\right)
\left(\frac{v_{\rm rec}}{1000\,\rm km\,s^{-1}}\right)^{5}
\,\rm erg\,s^{-1}.
  \label{Lkick}
\end{equation}

There is a radius outside of
which all particles in the disc will become unbound after the kick,
this is given by
\begin{equation}
R_{\rm ub}=\left( \cos \theta +\sqrt{\cos^2 \theta +1} \right)^2 R_{\rm rec}
\label{rub}
\end{equation}
\citep{Rossietal2010}, where $\theta$ is angle between the kick
velocity direction and the disc plane. This is maximal if the kick is
in the plane of the disc. We approximate this radius with
\begin{equation}
R_{\rm ub}=k R_{\rm rec}.
\end{equation}
We find the average parameter of $k=3.64$ by assuming an
  isotropic distribution for $\theta$ and integrating
  equation~(\ref{rub}) over $0<\theta<90^\circ$. This corresponds to
  $\theta=46.4^\circ$. We use this in our estimates for the mass
  $M_{\rm rec}$ in the rest of this work.

A further constraint on the circumbinary disc is that it must survive
until the black hole merger takes place. This depends sensitively on
the initial separation of the binary at the time when the black
  hole binary forms. The timescale for the decay of the orbit of an
equal mass binary as a result of gravitational waves is
\begin{equation}
t_{\rm GW}=1.1 \times 10^8
\left(\frac{a_{\rm b}}{10\,\rm R_\odot}\right)^4 
\left(\frac{M}{60\,\rm M_\odot}\right)^{-3}
\,\rm yr
\end{equation}
\citep{Peters1964}, where $a_{\rm b}$ is the separation of the binary.
The inspiral may be accelerated by the torque from a
prograde circumbinary disc \citep[e.g.][]{AN2005,Lodatoetal2009} and
even more significantly from a retrograde disc
\citep{Nixonetal2011a,Nixonetal2011b}.

The separation of the binary does not affect the structure of the
  disc at large radii from the binary. In this work we assume that the
  disc remains in quasi--steady state as the disc and the binary
  separation evolve.

\section{Accretion disc model}
\label{full}

We now consider the evolution of a circumbinary disc freely accreting
on to the binary components. The total mass of the disc is taken as
$M_{\rm d}=0.001\,M$. This otherwise arbitrary choice is designed to
illustrate that a detectable electromagnetic signal is possible even
from a residual disc of far lower mass than that lost ($\sim M$) in
the earlier evolution of the binary (see de Mink \& King 2017, Section
2). Since $M_{\rm d} \propto M_{\rm rec}$, equation (\ref{Lkick})
shows the effect of different choices.  We take a simple model in
which we assume that the disc aspect ratio $H/R$ is constant over the
radial extent of the disc.

Angular momentum transport in accretion discs is thought to be driven
by turbulence. In many cases, turbulence can be driven by the
magnetorotational instability \citep[MRI,][]{BH1991}. A critical level
of ionisation in the disc is required for this mechanism to operate:
if the disc is sufficiently hot, it may be sufficiently thermally
ionised. In parts of the disc with a lower temperature, it may be that
the surface layers of the disc are MRI--active because they are
ionised by external sources such as cosmic rays
\citep[e.g.][]{Glassgoldetal2004}.  A ``dead zone'' forms at the disc
midplane where the MRI does not operate
\citep{Gammie1996,Gammie1998,Perna2016}. In this Section we first
consider a disc model in which the MRI operates throughout, and then
we consider a disc model with a dead zone.

\subsection{Fully turbulent disc}

If the disc is fully MRI active, the lifetime of the disc is 
approximated by the viscous timescale at the circularisation radius of
the gas. The viscous timescale is
\begin{equation}
t_{\nu}=\frac{R^2}{\nu},
\label{visc}
\end{equation}
where the viscosity is parameterised with
\begin{equation}
\nu=\alpha c_{\rm s}H,
\end{equation}
with $\alpha$ is the dimensionless \cite{SS1973} viscosity
parameter. The scale height of the disc is $H=c_{\rm
  s}/\Omega$, where $c_{\rm s}$ is the sound speed, and so
\begin{equation}
\nu=\alpha\left(\frac{H}{R}\right)^2 R^2 \Omega
\end{equation}
and for typical values we find the viscous timescale to be
\begin{equation}
t_{\nu}=6.5\times 10^{5} \left(\frac{\alpha}{0.1}\right)^{-1} 
\left(\frac{H/R}{0.01}\right)^{-2} \left( \frac{M}{60\,\rm M_\odot}\right)^{-\frac{1}{2}} \left(\frac{R}{10^4\,\rm R_\odot}\right)^{\frac{3}{2}}
\,\rm yr.
\label{tnu}
\end{equation}
The lifetime of a fully MRI active disc with no binary torque is
relatively short, and in this case there may be no circumbinary material left
at the time of the merger unless the binary black holes are formed
with a separation $a_{\rm b}\lesssim 1\,\rm R_\odot$.

\subsection{Disc with a dead zone}

A dead zone forms at the disc midplane if two conditions are
satisfied. First, the temperature must be lower than the critical
required for the MRI to operate, $T_{\rm crit}\approx 800\,\rm K$
\citep{Umebayashi1988}. Above this temperature, the disc is thermally
ionised and MRI active. Second, the surface density must exceed
the critical that can be sufficiently ionised by external sources,
$\Sigma_{\rm crit}$. For a central black hole binary, the
only source of external ionisation is cosmic rays. These may ionise a
maximum surface density of $100\,\rm g\, cm^{-2}$
\citep[e.g.][]{Gammie1996, Matsumura2009,Zhuetal2009}. Effects such as
recombination may lead to a much smaller active layer
\citep[e.g.][]{Wardle1999,Balbus2001,Fleming2000,Martinetal2012a}. In
this work we take a fiducial value of $\Sigma_{\rm crit}=10\,\rm g\,
cm^{-2}$.

The inner edge of the dead zone is fixed by the radius at
which the temperature of the fully turbulent steady state disc model
is equal to the critical temperature required for the MRI.
The sound speed in the disc is given by
\begin{equation}
c_{\rm s}=\sqrt{\frac{kT}{\mu m_{\rm H}}},
\label{cs}
\end{equation}
where $k$ is the Boltzmann constant, $T$ is the temperature, $\mu=2.3$
is the mean molecular weight and $m_{\rm H}$ is the mass of a hydrogen
atom. Thus the disc temperature is
\begin{equation}
T=3.3\times 10^4
\left(\frac{H/R}{0.01}\right)^2
\left(\frac{M}{60\,\rm M_\odot}\right)
\left(\frac{R}{10\,\rm R_\odot}\right)^{-1}
\,\rm K.
\end{equation}
Setting $T=T_{\rm crit}$ we find the radius of the inner edge of the
dead zone as
\begin{equation}
R_{\rm dz}=417
\left(\frac{H/R}{0.01}\right)^{2} 
\left( \frac{M}{60\,\rm M_\odot}\right)
\left( \frac{T_{\rm crit}}{800\,\rm K}\right)^{-1}
\,\rm R_\odot.
\end{equation}
Note that both the temperature and the inner radius of the dead zone
depend sensitively upon the disc aspect ratio.

The surface density of the disc in the dead zone is higher than that
of a steady accretion disc since material builds up there over time
\citep[e.g.][]{MartinandLubow2011,MartinandLubow2013prop,Martinetal2012b}. The radius of
interest here, $R_{\rm ub}$, is likely too hot to be in a dead zone,
unless the disc aspect ratio is smaller than 0.01.  The dead zone acts
as a bottleneck in the accretion flow. We assume that the viscosity
in the dead zone is zero, although in practice there may be a small
amount of turbulence there
\citep[e.g.][]{MartinandLubow2013dza}. The surface density at radii
inside the dead zone radius is smaller than that of a disc without
a dead zone, and the reduced accretion on to the binary means that the
disc survives longer.  The bottleneck occurs at the innermost dead zone
radius, $R_{\rm dz}$. The accretion rate through the active surface
layer is approximately given by
\begin{equation}
\dot M_{\rm dz}=3 \pi \nu \Sigma_{\rm crit}.
\end{equation}
so
\begin{align}
\dot M_{\rm dz}=& \,\,7.2\times 10^{-9}
\left(\frac{\alpha}{0.1}\right)
\left(\frac{H/R}{0.01}\right)^{3} 
\left( \frac{M}{60\,\rm M_\odot}\right)\cr
& \times
\left( \frac{\Sigma_{\rm crit}}{10\,\rm g\, cm^{-2}}\right)
\left( \frac{T_{\rm crit}}{800\,\rm K}\right)^{-1/2}
\,\rm M_\odot \, yr^{-1}.
\end{align}
We estimate the lifetime of the disc as
\begin{align}
t_{\rm lifetime}=& \,\,\frac{M_{\rm d}}{\dot M_{\rm dz}},
\end{align}
where $M_{\rm d}$ is the total mass of the disc.
For our typical parameters this is
\begin{align}
t_{\rm lifetime} = & \,\, 8.4\times 10^{6}
\left(\frac{M_{\rm d}}{0.001\,M}\right)
\left( \frac{\alpha}{0.1}\right)^{-1}
\left(\frac{H/R}{0.01}\right)^{-3} 
\cr & \,\,\,\, \times 
\left( \frac{\Sigma_{\rm crit}}{10\,\rm g\, cm^{-2}}\right)^{-1}
\left( \frac{T_{\rm crit}}{800\,\rm K}\right)^{1/2}
\,\rm yr.
\end{align}
Again, this depends very sensitively on the disc aspect ratio. The
lifetime of this disc may be sufficiently long for material to remain
around until the merger.

Inside the inner dead zone radius, $R_{\rm dz}$, the disc behaves
as a steady accretion disc fed with accretion rate
$\dot M_{\rm dz}$. Away from the inner boundary, the steady state
surface density of the disc is approximately $\Sigma= \dot M_{\rm
  dz}/(3 \pi \nu)$. So if the recoil radius is smaller than than
the radius of the inner edge of the dead zone, the surface
density at the recoil radius is
\begin{align}
\Sigma(R_{\rm rec})=& \,\, 64.6\, 
\left( \frac{M}{60\,\rm M_\odot}\right)^{1/2}
\left( \frac{\Sigma_{\rm crit}}{10\,\rm g\, cm^{-2}}\right)
\left( \frac{T_{\rm crit}}{800\,\rm K}\right)^{-1/2}\cr
& \times
\left(\frac{H/R}{0.01}\right)
\left(\frac{R_{\rm rec}}{10\,\rm R_\odot}\right)^{-1/2}
\,\rm g\, cm^{-2}.
\label{sig}
\end{align}
This surface density is constant in time as long as the dead zone has
not been accreted.  The mass of the disc up to a radius $R_{\rm ub}$
is
\begin{align}
M_{\rm d,rec}=5.6\times 10^{-7}
\left( \frac{\Sigma_{\rm crit}}{10\,\rm g\, cm^{-2}}\right)
\left( \frac{T_{\rm crit}}{800\,\rm K}\right)^{-1/2}
\left(\frac{M}{60\,\rm M_\odot}\right)^2 \cr \times
\left(\frac{H/R}{0.01}\right) 
\left(\frac{v_{\rm rec}}{1000\,\rm km\,s^{-1}}\right)^{-3}
\left(\frac{k}{3.64}\right)^{3/2}
\,\rm M_\odot.
\label{mdrecacc}
\end{align}
This is quite small, but this is a
lower limit since if the inner edge of the dead zone is smaller than
$R_{\rm ub}$ the surface density must be higher than this.

In summary, a disc around a black hole that provides no tidal torque
on the disc is unlikely to survive until the time of the black hole
merger, unless it contains a dead zone. The dead zone significantly
extends the lifetime of the disc since material accretes on to the
binary at a much lower rate. However, the surface density in the inner
parts of the disc that feel the kick is low, and may not produce a
detectable signal at the expected source distances. Comparing
(\ref{mdrecacc}) with (\ref{Lkick}) suggests a peak luminosity of
$\sim 2 \times 10^{38}~{\rm erg\, s^{-1}}$

\section{Decretion disc model}
\label{zero}

We now consider the case of a binary torque strong enough to
prevent all accretion on to the binary. Angular momentum is added to
the inner parts of the disc. The disc spreads outwards while the
surface density decreases in time. We assume that the disc is fully
MRI active in this case.

\subsection{Quasi--steady state disc}

The disc evolves as a quasi--steady state decretion disc. The steady
state decretion disc solution satisfies
\begin{equation}
\nu \Sigma=\frac{(-\dot M)}{3\pi}\left[\left(\frac{R_{\rm out}}{R}\right)^{1/2}-1\right],
\end{equation}
where $R_{\rm out}$ is the outer disc edge and $-\dot M$ is the outward
accretion rate through the disc. Since we are interested in the inner
parts of the disc, we use the approximation $\nu \Sigma\propto
R^{-1/2}$ that is valid for $R \ll R_{\rm out}$.  We find the constant
of proportionality from the total mass of the disc that remains
constant in time.  We note that this approximation leads us to
underestimate the mass in the inner parts of the disc so our
calculated masses can be taken as a lower limit. We find the surface
density
\begin{align}
\Sigma = & \,\,3.94\times 10^4 
\left(\frac{M}{60\,\rm M_\odot}\right)
\left(\frac{M_{\rm d}}{0.001\, M}\right) 
\left(\frac{R}{10\,\rm R_\odot}\right)^{-1}
\left(\frac{R_{\rm out}}{10^4\,\rm R_\odot}\right)^{-1}\,\rm g\, cm^{-2}.
\end{align}
We note that the initial outer radius of the disc is determined
  by the angular momentum of the material that forms the disc. We show
  in Section~\ref{spread} that for late times, the initial outer
  radius does not significantly affect the properties of the disc.
The mass within radius $R_{\rm ub}$ is
\begin{align}
M_{\rm d,rec}=2.5\times 10^{-4}
\left(\frac{M}{60\,\rm M_\odot}\right)^2
\left(\frac{M_{\rm d}}{0.001\, M}\right)
\left(\frac{k}{3.64}\right)\cr \times
\left(\frac{R_{\rm out}}{10^4\,\rm R_\odot}\right)^{-1}
\left(\frac{v_{\rm rec}}{1000\,\rm km\,s^{-1}}\right)^{-2}
\,\rm M_\odot.
\label{mdrec2}
\end{align}
For our typical parameters, this is more than two orders of magnitude
larger than the accretion disc mass mass given in
equation~(\ref{mdrecacc}).  Comparing equation~(\ref{mdrec2}) with
(\ref{Lkick}) suggests a peak luminosity of $\sim 10^{41}~{\rm erg\,
  s^{-1}}$. However, since the decretion disc spreads outwards in
time, we must consider how this mass changes in time.

\subsection{Spreading of the disc}
\label{spread}

\begin{figure*}
\begin{centering}
\includegraphics[width=0.48\textwidth]{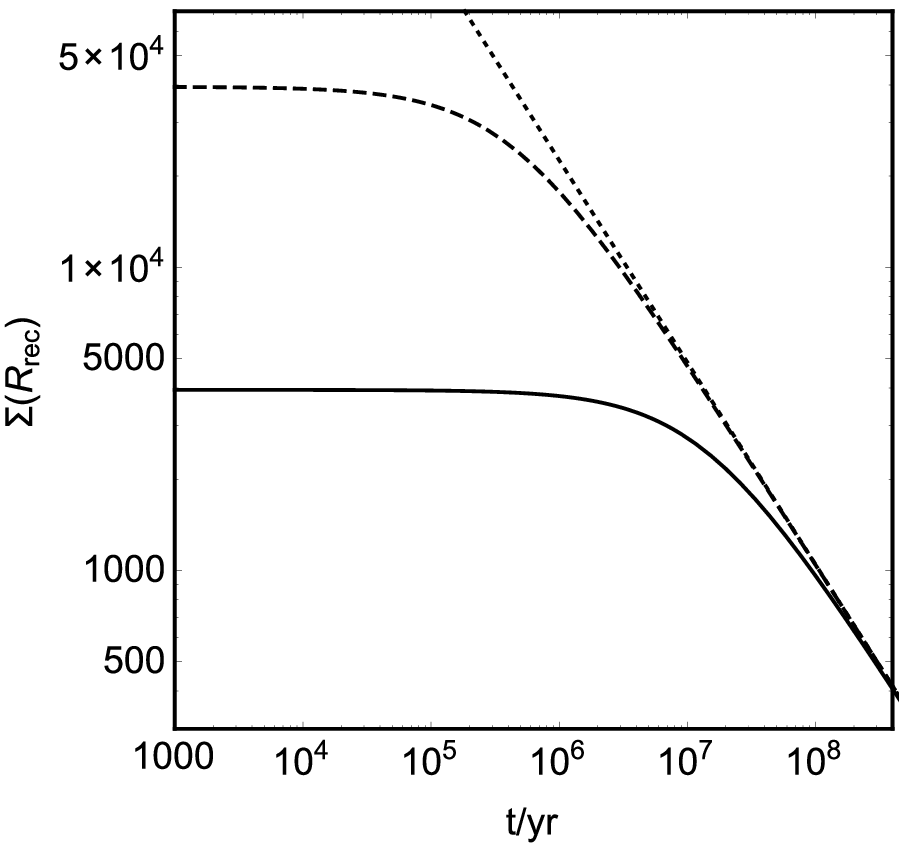}
\includegraphics[width=0.48\textwidth]{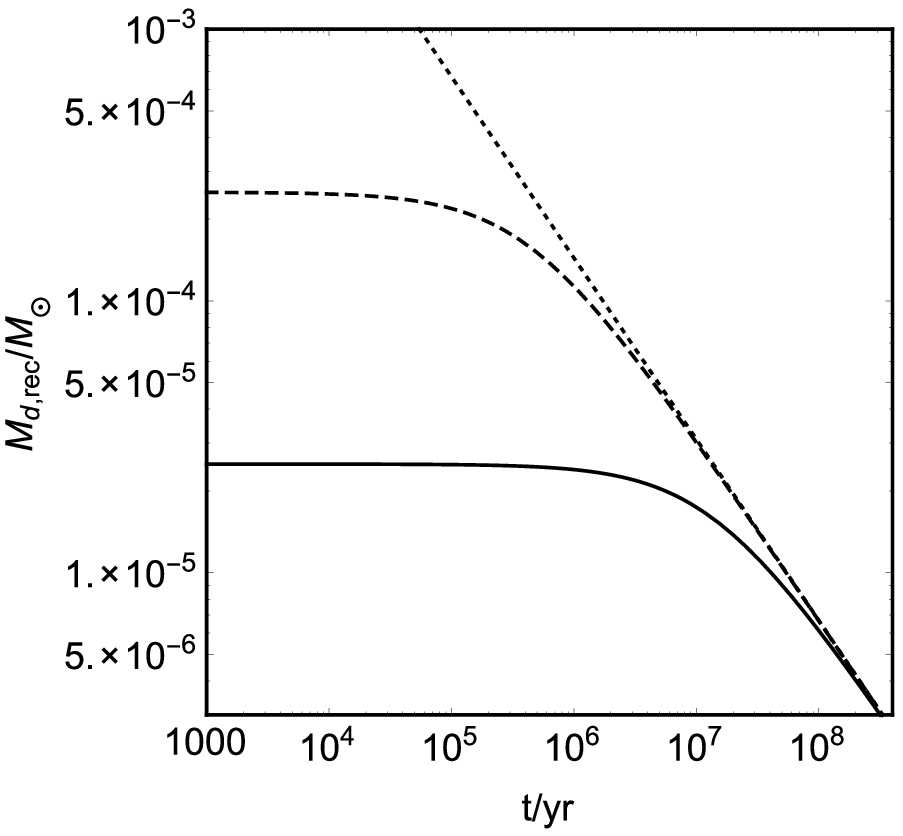}
\end{centering}
\caption{The decretion disc. Left: surface density at the recoil
  radius $R_{\rm rec}=10\,\rm R_\odot$ as a function of time. Right:
  mass of the disc in $R<R_{\rm ub}$. The initial outer edge of the
  disc is at $R_0=10^5\,\rm R_\odot$ (solid line) and $R_0=10^4\,\rm
  R_\odot$ (dashed line). The dotted lines show the analytic fits
  given in equations~(\ref{eqsig}) and~(\ref{eqm}).}
\label{rout}
\end{figure*}

The timescale for outward spreading of the disc is the viscous
timescale at its outer edge. We model this by assuming
that
\begin{equation}
\frac{d R_{\rm out}}{dt}=\frac{R_{\rm out}}{t_{\nu}(R_{\rm out})}.
\label{dr}
\end{equation}
We write the viscous timescale as
\begin{equation}
t_\nu=t_0 \left(\frac{R_{\rm out}}{R_0}\right)^{3/2},
\end{equation}
where $R_0$ is the initial outer disc radius and $t_0$ is the initial
viscous timescale at that radius, given by
equation~(\ref{visc}). Solving equation~(\ref{dr}) with
equation~(\ref{tnu}) we find that the outer radius of the disc evolves
as
\begin{equation}
R_{\rm out}=R_0\left(\frac{3}{2}\frac{t}{t_0}+1\right)^{2/3}.
\end{equation}
For an initial disc outer radius of $R_0=10^4\,\rm R_\odot$, the
viscous timescale there is $t_0=6.5\times 10^5\,\rm yr$ and for
$R_0=10^5\,\rm R_\odot$, the viscous timescale there is $t_0=2\times
10^7\,\rm yr$. The left panel of figure~\ref{rout} shows the surface
density of the disc at the recoil radius as a function of time for
initial outer disc radii $R_{\rm out}=10^4\,\rm R_\odot$ and $R_{\rm
  out}=10^5\,\rm R_\odot$ for our fiducial parameters $M=60\,\rm
M_\odot$, $M_d=0.001\,M$ and $R_{\rm rec}=10\,\rm R_\odot$.  The
surface density in the decretion disc is significantly higher than
that of the accretion disc with a dead zone, given in
equation~(\ref{sig}).  The right hand panel of Fig.~\ref{rout} shows
the mass of the disc in $R<R_{\rm ub}$.

For times greater than a few $t_0$, the outer disc edge is
approximately 
\begin{equation}
R_{\rm out}=R_0 \left(\frac{3}{2}\frac{t}{t_0}\right)^{2/3}.
\end{equation}
We can find the power law decay of the surface density and the mass of
the disc for late times with this expression. The surface density of
the disc at the recoil radius is
\begin{align}
\Sigma(R_{\rm rec})=& \,\,4.9\times 10^{3}
\left(\frac{M}{60\,\rm M_\odot}\right)^{2/3}
\left(\frac{M_{\rm d}}{0.001\, M}\right)
\left(\frac{\alpha}{0.1}\right)^{-2/3}
\left(\frac{H/R}{0.01}\right)^{-4/3}\cr & \times
\left(\frac{R_{\rm rec}}{10\,\rm R_\odot}\right)^{-1}
\left(\frac{t}{10^7\,\rm yr}\right)^{-2/3}
\,\rm g\,cm^{-2} .
\label{eqsig}
\end{align}
Further, the mass of the disc within $R_{\rm ub}$ 
is approximately
\begin{align}
M_{\rm d,rec}=& \,\,2.5\times 10^{-5}
\left(\frac{M}{60\,\rm M_\odot}\right)^{5/3}
\left(\frac{M_{\rm d}}{0.001\, M}\right)
\left(\frac{\alpha}{0.1}\right)^{-2/3}
\left(\frac{H/R}{0.01}\right)^{-4/3}\cr & \times
\left(\frac{k}{3.64}\right)
\left(\frac{v_{\rm rec}}{1000\,\rm km\,s^{-1}}\right)^{-2}
\left(\frac{t}{10^7\,\rm yr}\right)^{-2/3}
\,\rm M_\odot.
\label{eqm}
\end{align}
The dotted lines in Fig.~\ref{rout} confirm that these fits to the
surface density and the mass are good for times greater than
about a viscous timescale at the outer edge of the disc.

The mass in this decretion disc, for our fiducial parameters, is
larger than that of the accretion disc given in
equation~(\ref{mdrecacc}).  Further, if there is a dead zone in
the decretion disc, the surface density may be higher than that
predicted here for the fully turbulent decretion disc. The structure
would depend sensitively on the initial conditions. We will model this
scenario with time--dependent simulations in a future paper in order
to understand the disc evolution.

\begin{figure}
\begin{centering}
\includegraphics[width=0.48\textwidth]{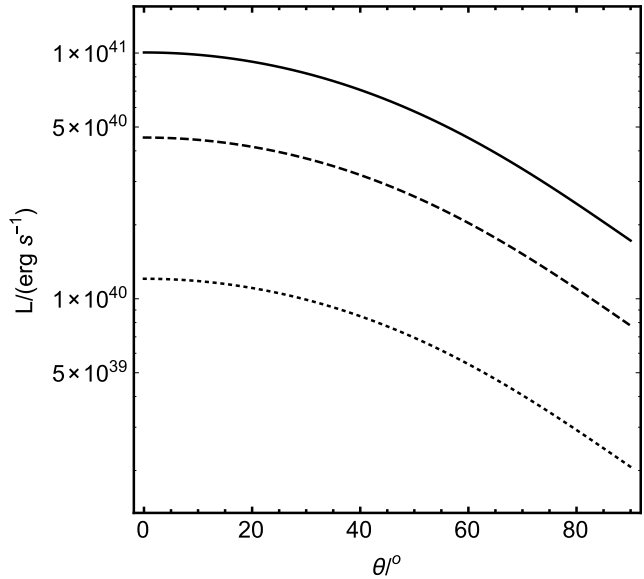}
\end{centering}
\caption{The luminosity of the decretion disc. The initial outer edge
  of the disc is at $R_0=10^4\,\rm R_\odot$. The times shown are $t=0$
  (solid line), $t=10^6\,\rm yr$ (dashed line) and $t=10^7\,\rm yr$
  (dotted line). }
\label{luminosity}
\end{figure}

Fig.~\ref{luminosity} shows the luminosity of the decretion disc as a
function of the angle between the velocity kick and the disc plane,
$\theta$, as calculated by equation~(\ref{Lkick}). We include the
$\theta$ dependence from equation~(\ref{rub}). The luminosity is
maximal for a kick that is in the plane and decreases by almost an
order of magnitude for a kick that is perpendicular to the disc.

\section{Conclusions}
\label{conc}

We have explored the evolution of a circumbinary disc around a merging
black hole system in the two extreme limits of an accretion disc and a
decretion disc. The accretion disc feels no torque from the
binary. The viscous timescale of such a disc is short and it is
unlikely that significant circumbinary material remains at the time of
the merger unless there is a dead zone in the disc. The dead zone
restricts the accretion flow through the disc and the disc life time
is significantly extended. However, the surface density of the disc
around the recoil radius is low, unless the dead zone extends into
this region. 

In the opposite limit, where a strong binary torque prevents
accretion, the disc behaves as a decretion disc.  The surface density
at the recoil radius is much larger than in the accretion disc, even
without a dead zone. 

The physical conditions in a disc around a BH--BH binary probably conform
to the case of a decretion disc, held out by tidal torques from the binary, and
containing dead zones. As we have seen, there is significant mass close to 
the recoil radius in this case. We conclude that
dynamical readjustment of the disc after the BH merger
is likely to release significant energy in electromagnetic form. Since all
of the disc matter, including its outer skin, is shocked simultaneously, it appears
unlikely that this energy is significantly trapped within the shocked disc. The
prompt appearance of an electromagnetic counterpart, delayed by the light--travel 
time to the recoil radius (a few hours) therefore seems promising \citep[cf][]{deMink2017}.

\section*{Acknowledgments} 

RGM acknowledges the hospitality of Leicester University during a
visit where parts of this work were completed. RGM acknowledges
support from NASA through grant NNX17AB96G. CJN is supported by the
Science and Technology Facilities Council (grant number
ST/M005917/1). FGX is supported in part by National Program on Key
Research and Development Project of China (grant Nos. 2016YFA0400804),
the Youth Innovation Promotion Association of CAS (id. 2016243), and
the Natural Science Foundation of Shanghai (grant No. 17ZR1435800). The
Theoretical Astrophysics Group at the University of Leicester is
supported by an STFC Consolidated Grant.

\bibliographystyle{mn2e} 

\label{lastpage}
\end{document}